\documentclass{ws-mpla}

\newcommand{\lan}{\langle}
\newcommand{\ran}{\rangle}
\newcommand{\non}{\nonumber}

\def\bea{\begin{eqnarray}}
\def\ena{\end{eqnarray}}

\usepackage{amssymb}\usepackage{amsmath}

\def\n{{\bf n}}

\def\U{{\bf U}}

\def\muK{{\rm \mu K}}

\def\bi{\begin{itemize}}
\def\ei{\end{itemize}}

\def\n2{N$_{2}$}
\def\4he{$^{4}$He}
\def\cm3{cm$^3$}

\def\beq{\begin{equation}}
\def\eeq{\end{equation}}

\usepackage{srcltx}
\usepackage{srctex}
\begin{document}

\title{The Polarization of the Cosmic Microwave Background Due to Primordial Gravitational Waves}

\author{\footnotesize Brian G. Keating}

%\footnote{
%University of California, San Diego, Mail Code 0424, La Jolla, CA,
%92093-0424, United States of America}}

\address{University of California, San Diego, Mail Code 0424, La Jolla, CA,
92093-0424, United States of America \\ bkeating@ucsd.edu }

\author{\footnotesize Alexander G. Polnarev}

\address{Queen Mary and Westfield College, University of London, Mile End Road, London, United Kingdom \\ agp@maths.qmw.ac.uk}

\author{\footnotesize Nathan J. Miller}

%\footnote{
%University of California, San Diego, Mail Code 0424, La Jolla, CA,
%92093-0424, United States of America}}

\address{University of California, San Diego, Mail Code 0424, La Jolla, CA,
92093-0424, United States of America \\
nmiller@physics.ucsd.edu}

\author{\footnotesize Deepak Baskaran}
%\footnote{
%Cardiff University, Cardiff, Wales, United Kingdom}}

\address{Cardiff University, Cardiff, Wales, United Kingdom \\ Deepak.Baskaran@astro.cf.ac.uk}

%\footnote{
%Queen Mary and Westfield College, University of London, Mile End
%Road, London, United Kingdom}}

\maketitle

\pub{Received (Day Month Year)}{Revised (Day Month Year)}

\begin{abstract}
We review current observational constraints on the polarization of
the Cosmic Microwave Background (CMB), with a particular emphasis
on detecting the signature of primordial gravitational waves. We
present an analytic solution to the Polanarev approximation for
CMB polarization produced by primordial gravitational waves. This
simplifies the calculation of the curl, or B-mode power spectrum
associated with gravitational waves during the epoch of
cosmological inflation. We compare our analytic method to existing
numerical methods and also make predictions for the sensitivity of
upcoming CMB polarization observations to the inflationary
gravitational wave background. We show that upcoming experiments
should be able either detect the relic gravitational wave
background or completely rule out whole classes of inflationary
models.

\end{abstract}

\ccode{PACS Nos.: include PACS Nos.}

\markboth{Brian G. Keating, Alexander G. Polnarev, Nathan Miller,
and Deepak Baskaran} {Gravitational Waves and the Polarization of
the CMB}

%%%%%%%%%%%%%%%%%%%%% Publisher's Area please ignore %%%%%%%%%%%%%%%
%
%\catchline{}{}{}{}{}
%
%%%%%%%%%%%%%%%%%%%%%%%%%%%%%%%%%%%%%%%%%%%%%%%%%%%%%%%%%%%%%%%%%%%%

\section{Introduction}
The cosmic microwave background (CMB) is one of the most powerful
and precise cosmological probes. Because the CMB photons we
observe today probe the physics of the early universe during the
epoch of linear gravity, the CMB is often referred to as a
``snapshot'' of the primordial universe. The CMB has the promise
to address the most fundamental cosmological questions: the
geometry and age of the universe, the matter-energy content of the
universe, the ionization history and the spectrum of primordial
perturbations.

This review addresses the theoretical foundations of CMB
polarization generated by cosmological gravitational waves with
particular emphasis given to analytic and numerical results which
encode its behavior most cogently. Gravitational waves, in
contrast to adiabatic perturbations (which are the dominant source
of CMB temperature and polarization anisotropy) imprint a unique
divergence-free pattern of polarization on the
sky\cite{polnarev85,seljak1997}. This pattern is called
``B-mode''\cite{Zalandseljak97} or ``curl-mode''\cite{kamkossteb}
polarization. Although there may be a substantial GWB contribution
to CMB temperature anisotropy, its effect on the CMB temperature
is nearly completely degenerate with other cosmological
parameters\cite{kamionkowski97,selzal97}. This is unsurprising
since the CMB temperature is a scalar quantity. However, the
tensorial nature of CMB polarization permits separation of scalar
fluctuations from tensor, GWB-generated, fluctuations. CMB
polarization maps can be decomposed into two
terms\cite{kamionkowski97,selzal97}. One term is the gradient of a
scalar potential and is invariant under parity transforms (often
called ``E-mode" in analogy to the electric field/scalar
potential). The second component is the curl of a vector potential
(``B-mode"). Scalar perturbations have no handedness so the
primary CMB curl-mode component exists only if there is a GWB.
Besides the obvious importance of a new method of (indirect)
detection of gravitational waves, detection of the B-mode signal
provides the cleanest, and perhaps only window into unique
predictions of the inflationary cosmological paradigm.

Numerical methods for simulating CMB temperature and polarization
power spectra have revolutionized cosmology
\cite{SeljakZal1996,Lewis:2002ah}. Without such codes parameter
estimation from CMB data sets would be difficult, if not
impossible. However, the complexity of these codes has hidden the
underlying physics. For this reason there is considerable interest
in analytical approaches for calculating CMB polarization caused
by cosmological gravitational waves (see for example
Ref.~\refcite{polnarev85,basko,ColeFrewinPolnarev95,keating1998,pritchardkamionkowski2005,zhaozhang2005}).

This review elucidates an analytical approach to the problem using
the fundamental physical characteristics of gravitational waves to
significantly simplify calculations. Our method simplifies the
comparison between theoretical predictions and future
observational results. We compare our results to existing
numerical methods and summarize current observational results. We
conclude with by making predictions for ground, balloon, and
space-based observations in the upcoming decade.

\section{Inflation and Primordial Gravitational Waves}

Inflation\cite{guth81} is a bold cosmological paradigm which has
provided key insights into many observations of modern cosmology.
Inflation provides an explanation for the observed
spatial-flatness of the universe, motivates the lack of
topological defects such as magnetic monopoles, and explains the
exquisite isotropy of the cosmic microwave background (CMB) while
simultaneously providing a mechanism to generate the observed
fluctuations. However, along with its many successes has come
increased scrutiny. Many of inflation's key predictions are shared
by alternative models. While skeptics\cite{khoury01,magueijo03}
are increasingly challenged to attack it, inflation's proponents
can only claim circumstantial evidence in its favor. Only new
discoveries will provide the data required to break the deadlock.
A conclusive detection of the primordial gravitational wave
background (GWB) predicted by inflation would be ``the smoking
gun" confirming the inflationary model beyond a reasonable doubt.
No other known cosmological mechanism mimics the imprint of the
GWB on the polarization of the CMB.

One of inflation's successful predictions is its solution to the
``horizon problem" - the observation that regions of the universe
share the same thermodynamic temperature despite never being in
thermal, or even causal, contact. Inflation solves the horizon
problem via a superluminal expansion of the universe at very early
times, prior to the ordinary Hubble expansion observed today. This
rendered the entire observable universe within causal contact
initially. This expansion also accounts for the (seemingly)
finely-tuned spatial flatness of the universe observed by CMB
temperature anisotropy
experiments\cite{deBernardis00,balbi00,pryke02,sievers03,spergel2003,goldstein03}.

This review focuses on inflationary generated gravitational wave,
or \emph{tensor} perturbations. However, inflation also predicts a
nearly scale-invariant spectrum of energy density, or scalar,
perturbations. That any perturbations remain after the universe
expanded by 60 \textit{e}-folds is astonishing! Yet, surprisingly,
the observed fluctuation level arises naturally as magnified
quantum fluctuations of the scalar field that drove inflation (the
inflaton) (see [\refcite{liddlelyth2000}] for a review of
inflationary perturbation theory). Following the inflationary
epoch, the size of the residual fluctuations were imprinted on the
surface of last scattering and are observable in the CMB. The
observed size of the fluctuations, e.g., [\refcite{hinshaw1996}],
and the correlations\cite{spergelzaldarriaga1997,peiris2003}
between the CMB's temperature and polarization patterns at
super-horizon scales motivates, but does not prove, the
inflationary paradigm.

Inflation not only produces these perturbations, but also endows
them with synchronized initial phases. This is crucial as it
allows the perturbations to grow without cancellation, as would
occur if the initial phases were uncorrelated. WMAP and other CMB
experiments, in combination with large-scale-structure
observations\cite{eisenstein05,cole05}, probe the scalar
perturbation spectrum and are consistent with inflation.

Regrettably, neither flatness nor smoothness are unique to
inflation. Both have long histories in cosmology. The flatness of
the universe was anticipated prior to inflation on quasi-anthropic
grounds\cite{dicke65}. A scale-invariant primordial matter/energy
perturbation spectrum (i.e., deviations from perfect smoothness)
was also predicted well before
inflation\cite{harrison70,peebles70,zeldovich72}, though no
mechanism to produce the perturbations was given in these early
works. These perturbations, combined with the universe's spatial
flatness, increase the plausibility of the inflationary paradigm,
since the amount of expansion required for flatness should have
also smoothed any initial perturbations to zero.

Two more observations are noteworthy for their consistency with
the inflationary paradigm: the low abundance of
relicts\cite{borner03} (e.g., no magnetic monopoles) and the
observed Gaussianity of perturbations\cite{komatsu03}. Since, as
many authors have pointed out\cite{turner02,penrose05}, inflation
has passed so many observational tests any replacement theory
would need to look very much like inflation. However, while there
is abundant circumstantial evidence, there is one unique
prediction of inflation that cannot be mimicked: a primordial
gravitational wave background; thereby cementing its status as
``the smoking gun of inflation"\cite{kamkosowsky99}.

Inflation posits a new scalar field (the inflaton) and specifies
its action potential, and thereby, its dynamics (calculated using
the standard methods of scalar field theory\cite{borner03}). All
theories of inflation produce a GWB, though some at unobservable
levels. All are scalar field theories, incorporating fluctuations
via quantum perturbation theory. While the identity of the
inflaton is unknown, specifying the inflaton potential has
immediate observational consequences\cite{kinney97}. The most
important inflationary measure of the GWB is the tensor-to-scalar
ratio, $r$, since it parameterizes the unique prediction of
inflation. A detection of $r$ would simultaneously reveal both the
epoch of inflation and its energy scale\cite{turner02}. If, as
theorists speculate, inflation is related to the GUT-scale (grand
unified theory) then detection of the B-mode signature would probe
physics at at the $10^{16}$ GeV scale\cite{kamkosowsky99}.

\section{CMB Polarization}

The CMB has been the most effective tool to appraise inflation
because the CMB is the earliest electromagnetic ``snapshot" of the
universe, and as such, probes the universe in a particularly
pristine state - before gravitational and electromagnetic
processing. Since gravity is the weakest of the four fundamental
forces, gravitational radiation (the inflationary GWB) probes much
farther back - to $\sim 10^{-38}$ sec after the Big Bang.

CMB polarization is generated by both scalar and tensor
perturbations. The two are related in all models of inflation
since both are generated by fluctuations of the same quantum
field. Inflation may have occurred at energies too low to detect
(or not at all) but the scalar-tensor relation provides a powerful
consistency check\cite{liddlelyth2000} - insurance against a
false-positive claim.

All inflationary observables are determined by the inflaton
potential. The process is invertible, affording the opportunity to
reconstruct, or significantly constrain, the inflaton potential
from CMB measurements of the GWB. Assuming the inflationary
paradigm is correct, this provides a window into physics at energy
scales below $10^{16}$ GeV even if future CMB observations produce
a null result. Without observational constraints, the number of
potential inflationary models has swelled to daunting
levels\cite{liddlelyth2000,kinney2003}. Upcoming experiments
should be able to rule out many of these models, even with a null
result as we discuss in section
\ref{section:inflation_predictions}.

CMB temperature anisotropy measurements can determine the
amplitude of the primordial power spectrum, $A_s$, and the scalar
power-spectral index, $n_s$, far better than CMB polarimeters but
only polarization sensitive experiments can measure the
key-prediction of inflation -- the tensor-to-scalar ratio, $r$, if
it is below $r \simeq 0.3$, which is near the current $2\sigma$ upper
limits as discussed in \S\ref{sec:results}. A joint detection of
$A_s$, $n_s$, and $r$ will serve to reconstruct the inflaton
potential (as well as completely rule out the ekpyrotic and
variable-speed-of-light models). This observation requires an
understanding of the effects of gravitational waves on the CMB, to
which we now turn.

\section{Primordial Gravitational Wave CMB Perturbations}

Gravitational waves perturb the metric tensor describing the
geometry of the early universe. The general, perturbed, metric of
a flat FLRW universe can be written as

\bea ds^2 = a^2\left(\eta\right)\left[ -d\eta^2 +
\left(\delta_{ij} + h_{ij}\right)dx^idx^j\right]. \ena

The components $h_{ij}\left(\eta,{\bf x}\right)$ represent the
gravitational waves and can be expanded into spatial Fourier
harmonics $e^{\pm i {\bf k\cdot x}}$,

\bea h_{ij}\left(\eta,{\bf x}\right) =
\frac{1}{(2\pi)^{3/2}}\int~d{\bf
k}\sum_{s=1,2}\stackrel{s}{p}_{ij}\left({\bf
k}\right)\left[{h}_k\left(\eta\right)e^{i{\bf k\cdot
x}}\stackrel{s}{c}_{\bf k}+ {h}_k^*\left(\eta\right)e^{-i{\bf
k\cdot x}}\stackrel{s}{c}_{\bf k}^*\right],\label{fourierh}\ena

where ${\bf k}$ is a time-independent wave vector, and
$k=(\delta_{ij}k^ik^j)^{1/2}$. The wave number $k$ defines the
wavelength measured in units of laboratory standards by $\lambda =
2\pi a /k$. The polarization tensors $\stackrel{s}{p}_{ij}({\bf
k})$ have different forms depending on whether they represent
gravitational waves, rotational perturbations, or density
perturbations\cite{lpg&sb2002}. In the case of gravitational waves
the polarization tensors can be expressed in terms of two mutually
orthogonal unit-vectors $({\bf l},{\bf m})$ lying in the plane of
the wave front (i.e. perpendicular to ${\bf k}$), \bea
\stackrel{1}{p}_{ij}=l_il_j-m_im_j,~~\stackrel{2}{p}_{ij}=l_im_j+l_jm_i,\ena
and obey the conditions \bea
\stackrel{s}{p}_{ij}\delta^{ij}=0,~~\stackrel{s}{p}_{ij}k^{i}=0,\nonumber
\\ \stackrel{s'}{p}_{ij}({\bf k})\stackrel{s}{p}^{ij}({\bf k})=2\delta_{s's}.\ena

For a classical gravitational wave field, the quantities
$\stackrel{s}{c}_{\bf k}$ and $\stackrel{s}{c}_{\bf k}^\dag$ in
Eq. \ref{fourierh} are arbitrary complex (conjugate) numbers.

The Fourier expansion allows us to reduce the problem of evolution
of the perturbed gravitational field to the evolution of mode
functions ${h}_k(\eta)$ for each individual mode $k$. For a single
Fourier component the amplitude of the primordial gravitational
wave obeys the following wave equation\cite{dodelson2003}:

\beq \frac{d^2{h}_k}{d\eta^2}+\frac{2}{a}
\frac{da}{d\eta}\frac{d{h}_k}{d\eta}+ k^2{h}_k=0,
\label{eq:gwh}\eeq

This equation ignores the damping effects of anisotropic stress
provided by cosmological neutrinos\cite{weinberg2004}. For
$k\eta\ll1$ (wavelength larger than the cosmological horizon) the
last term in the above equation can be dropped out, and the
amplitude of the gravitational wave is ``frozen" (${h}_k\sim
const$). For $k\eta\gg1$ (wavelength smaller than the horizon) the
solutions for ${h}_k$ are damped plane waves. Since larger $k$
perturbations enter the horizon earlier, their contribution to the
tensor spectrum is more heavily damped.

Equation \ref{eq:gwh} allows us to study the gravitational wave
perturbations in two cosmological epochs: 1) in the matter
dominated epoch, governed by matter with effective equation of
state $p=0$, where the scale factor behaves as
$a(\eta)\propto\eta^2$ and 2)in the radiation dominated epoch,
governed by the effective equation of state $p=\epsilon/3$, with
$a(\eta)\propto\eta$.

In the general case with cosmic scale factor $a(\eta)$, Eq.
\ref{eq:gwh} might not allow for an analytical solution. For a
mixed matter-radiation Universe, the scale factor is given by the
expression \cite{Chernin1966}

\bea a(\eta) = \left[ \frac{4l_H}{1 + \sqrt{1+4\eta_*^2}} \right]
\eta(\eta + 2\eta_{*}),\label{scale-m-r}\ena

where $\eta_* = (\sqrt{2}-1)\eta_{eq},$ and $\eta_{eq}$ is the
time of matter-radiation equality. The participating constants
have been chosen such that the value of the scale factor at the
present epoch $\eta_0$ is $a(\eta_0) = 2l_H$, which leads to the
value of the time at present epoch $\eta_0  = \left[(1-2\eta_{*})
+ \sqrt{1+4\eta_{*}^2}\right]/2 $. With such a convention, the
wave whose wavelength, $\lambda$, today is equal to present Hubble
radius, carries the constant wavenumber $k_H= 4\pi$. The scale
factor has two asymptotic regions: $\eta\ll\eta_*$ corresponding
to the radiation dominated epoch with $a(\eta)\propto\eta$, and
$\eta\gg\eta_*$ corresponding to the matter dominated epoch with
$a(\eta)\propto\eta^2$.

An elegant method to analyze the evolution of gravitational waves
in a such a universe is to smoothly approximate the above scale
factor with power-law scale factors, for which we can analytically
solve the  equation (\ref{eq:gwh}). The scale factor (Eq.
\ref{scale-m-r}) is well approximated by
\bea a(\eta) = a_0\cdot\eta, ~~~\eta\leq\eta_{**}, \nonumber \\
a(\eta) = 2l_H\cdot\left(\eta
-\eta_m\right)^2,~~~\eta_{**}\leq\eta.\label{jointscalefactor}\ena
Imposing continuity of $a(\eta)$ and $a'(\eta)$ at
matter-radiation equality, $\eta_{**}$, fully determines all
participating constants. \bea \eta_{**} = -\eta_m =
\frac{1}{2\sqrt{1+z_{eq}}},\nonumber \\ = a_0
\frac{4l_H}{\sqrt{1+z_{eq}}}.\ena The general solution to Eq.
\ref{eq:gwh} for the scale factor (Eq. \ref{jointscalefactor}) is
given by \bea h_k(\eta) = \left\{\begin{array}{l}
\frac{1}{a_0\eta}\left(C_ke^{-ik\eta}
+ D_ke^{ik\eta}\right),~~~\eta<\eta_{**}\\
\\ \frac{\sqrt{k}}{2l_H(\eta-\eta_m)^{3/2}}\left[A_kJ_{3/2}\left(\frac{}{}k(\eta - \eta_m)\right)-iB_kJ_{-3/2}
\left(\frac{}{}k(\eta - \eta_m)\right)\right],~~~\eta_{**}<\eta
\end{array}\right.\label{hm-r}\ena

The constants $C_k$ and $D_k$ are determined by the evolution of
the gravitational waves before radiation domination. Since the
radiation dominated era is believed to have been preceded by
inflation the values of the coefficients $C_k$ and $D_k$, which
give the spectral characteristics of the gravitational wave field,
are determined by the physics of inflation and initial conditions
(see for example \cite{Grishchuk93,allenkoranda1994}). It follows
that due to the initial stage of rapid expansion $ C_k \approx
-D_k$.

The fact that $C_k\approx -D_k$ shows that the gravitational wave
modes, ${\bf k}$, are (almost) standing waves at the radiation
epoch. To find the coefficients $A_k$ and $B_k$ in the solution
(\ref{hm-r}) $h_k(\eta)$ and $dh_k(\eta)/d\eta$ must be joined
continuously at the transition point $\eta = \eta_{**}$
\cite{NgSteliopolis1994}.

The above analytical approximation (\ref{hm-r}) sufficiently well
approximates the solution to the  equation (\ref{eq:gwh}) in the
case of the scale factor given by (\ref{scale-m-r}). Numerical
calculations show that the above analytical approximations work
very well for wavenumbers $k$ satisfying $k\Delta\eta_{eq}<1$,
where $\Delta\eta_{eq}$ is the characteristic time scale of change
from the radiation-dominated era to matter-dominated
era\cite{allenkoranda1994}.

%%%%%%%%%%%%%%%%%%%%%%%%%%%%%%%%%%%%%%%%%%%%%%%%%%%%%%%%%%%%%%%%%%%%%%%%%%%%%%%%%%%%%%%%%%%%%%

\section{CMB Polarization Observables}

Having considered the underlying cosmological behavior of tensor
perturbations we now turn to the imprint of tensor gravitational
waves on CMB polarization using the equation of radiative
transport. To begin we consider a polarized electromagnetic wave
with angular frequency, $\omega$:
$${\bf E} = E_{y0}\sin (\omega t - \delta _y){\bf \hat{y}} +
E_{x0}\sin (\omega t - \delta _x){\bf \hat{x}}.$$ The polarization
state of is characterized by the Stokes parameters: $I, Q, U,$ and
$V$:
$$I = I_y + I_x ,$$ with $I_y = \lan E_{y0}^2\ran$ and $ I_x =
\lan E_{x0}^2\ran$. $I$ is the total intensity of the radiation,
and is always positive. The other Stokes parameters are defined as

\begin{eqnarray}
Q &=& I_y - I_x \rm{\;and}\non\\
U &=& 2E_{y0}E_{x0}\cos (\delta _y - \delta_x)\non \\
V &=& 2E_{y0}E_{x0}\sin (\delta _y - \delta_x)\non
\end{eqnarray} where $Q$ and $U$ quantify the linear polarization of the wave, and
$V$ quantifies the degree of circular polarization. The
polarization fraction is $\Pi = \frac{\sqrt{Q^2 + U^2 +
V^2}}{I}$, and the polarized intensity is $I_{{\rm pol}} \equiv
\Pi \times I$.
The Stokes parameters comprise a symbolic vector ${\bf \hat{I}}$
introduced by Chandrasekhar\cite{chandrasekhar60} and related to the Stokes
parameters in the following way:

$$
{\bf \hat{I}} = \left( \begin{array}{c} I_{x} \\ I_{y} \\ U \\ V \end{array} \right) .$$

Thomson scattering only produces linear CMB polarization, implying $V=0$, so we will only consider the
symbolic 3-vector: ${\bf \hat{I}} = \left( \begin{array}{c} I_x \\ I_y \\ U
\end{array} \right).$

Polarized radiation in the presence of cosmological metric
perturbations is represented as state vector describing the
occupation numbers of polarized radiation\cite{basko,polnarev85}
\bea{\bf \hat{n}} = \frac{c^2}{h \nu^3}{\bf \hat{I}}={\bf
\hat{n}_0 }+n_0{\bf \delta\hat{n}
}=n_0\left[\left(\begin{array}{c}1\\1\\0\end{array}\right)+{\bf
\delta \hat{n}}\right].\label{n+dn}\ena

The Boltzmann equation of radiative transfer written in terms of
${\bf \hat{n}}(\eta,x^{\alpha},\theta,\psi)$ is:

\beq \frac{ \partial {\bf \hat{n}}}{\partial \eta} + {\bf
e^{\alpha}} \cdot \frac {\partial {\bf \hat{n}}}{\partial
x^{\alpha}} = -\frac{\partial {\bf \hat{n}}}{\partial \nu}
\frac{\partial {\nu}}{\partial \eta} - q( {\bf \hat{n}} - {\bf
\hat{J}}) \label{eqert} \eeq and \beq {\bf  \hat{J}} = \frac{1}{4\pi}\int_{0}^{\pi}\int^{2\pi}_{0}  {\bf \hat{\hat{P}}}
(\theta,\psi,\theta^{\prime},\psi^{\prime}) {\bf \hat{n}}
(\eta,x^{\alpha},\nu,\theta^{\prime},\psi^{\prime})\sin{\theta^{\prime}}d\theta^{\prime}
d\psi^{\prime}, \eeq where $q = \sigma_T N_e a$, $a$ is the cosmological
scale factor, ${\bf  \hat{J}}$ is the ``scattering'' or
``collisional'' term which is a function of the angular variables
only, with primed angular variables corresponding to the photon
direction before scattering and unprimed angular variables
corresponding to the photon direction after scattering. ${\bf
\hat{\hat{P}}}$ is the scattering matrix\cite{chandrasekhar60}, $\sigma _T$ is the Thomson cross section,
and $N_e$ is the comoving number density of free electrons.  The
coupling of the gravitational waves to the radiation is manifested
in the first term on the right side of Eq. \ref{eqert}:
$$\frac{1}{\nu}\frac{d{\nu}}{d\eta}= \frac{1}{2} \frac{\partial h_{\alpha\beta}}{\partial \eta}
e^{\alpha}e^{\beta}.$$

Let us introduce ${\bf\hat{n}_0} = n_0 \left(
\begin{array}{c} 1
\\ 1 \\ 0
\end{array} \right)$ corresponding to unpolarized isotropic thermal radiation (zero-th order
approximation), with $~{n_0= [\exp{h\nu/k_b T - 1}]^{-1}}$, which
depends only on the photon frequency $\nu$ and corresponds to the
Planck spectrum.

The angular symmetry of ${\bf \hat{\hat{P}}}$ requires: \beq {\bf
\hat{J}}(\hat{\bf n}_0) = \frac{1}{4\pi}\int^{\pi}_{0}\int^{2\pi}_{0}{\bf \hat{\hat{P}}}
(\theta,\psi,\theta^{\prime},\psi^{\prime}){\bf
\hat{n}_0}\sin{\theta^{\prime}}d\theta^{\prime}
d\psi^{\prime}={\bf \hat{n}_0}. \label{Jn_o}\eeq It should be
mentioned that the form of the scattering integral in (\ref{Jn_o})
assumes that the chosen reference frame is comoving with the
scattering electrons. In the case of scalar perturbations an
additional Doppler term may arise due to the movement of electrons
with respect to the chosen reference frame\cite{Bond1984}. In the
case of gravitational waves this Doppler term does not arise,
since it is always possible to choose our synchronous reference
frame to be comoving with the scattering electrons.

In Eq. \ref{n+dn}, ${\bf \delta\hat{n}}$ is the first order
correction to the uniform, isotropic, and unpolarized radiation
described by ${\bf \hat{n}_0}$. This perturbation is comprised of
an (unpolarized) term due to angular anisotropy of the photon
distribution, ${\bf \hat{n}_A}$, and a polarized term, ${\bf
\hat{n}_{\Pi}}$, so ${\bf \delta\hat{n}} = {\bf \hat{n}_A}+ {\bf
\hat{n}_{\Pi}}$.

The anisotropic and polarized components are functions of
conformal time, $\eta$, co-moving spatial coordinates,
$x^{\alpha}$, photon frequency, $\nu$, and photon propagation
direction specified by the unit vector ${e}^{\alpha}(\theta,\phi)$ with
polar angle, $\theta$, and azimuthal angle, $\phi$. All of the
polarization-specific cosmological phenomena discussed in this
review can be described by their effect on ${\bf \delta\hat{n}}$.
These effects, if measurable, can be used to evaluate and refine
the standard cosmological model. Since CMB polarization depends on
anisotropy, it probes \emph{all} the same underlying physics and,
in addition, several cosmological effects are \emph{only}
observable via CMB polarization and not CMB anisotropy. For this
reason, CMB polarization is a valuable cosmological tool. The
penalty we pay, however, is that the Thomson scattering which
produces CMB polarization has a fairly inefficient coupling to
cosmological perturbations. We will evaluate the cosmological
phenomena to which CMB polarization is sensitive in subsequent
sections.

We retain only the zero and first order perturbation terms in
$h_{\alpha\beta}$. Since $ \frac{d{\nu}}{d\eta}$ is of the first
order, we can replace $\frac{d{\bf \hat{n}}}{d\nu}$ by $
\frac{d{\bf \hat{n}_0}}{d\nu_0}$ in Eq. \ref{eqert} ($\nu_0$ is
the unperturbed frequency). This implies that the frequency
dependence of both the polarization and anisotropy is given by the
same factor:\cite{basko} $$ \gamma \equiv \frac{\nu_0}{n_0}
\frac{dn_0}{d\nu_0}.$$

We now concentrate on the first-order terms. In the following we will identify ${\bf \hat{J}}({\bf \hat{n}})={\bf \hat{J}_1}({\bf
\delta\hat{n}})$. An arbitrary gravitational wave can be
considered as a linear superposition of plane gravitational waves.
Due to the linear nature of the problem the anisotropy and
polarization generated by an arbitrary gravitational wave is the
linear superposition of anisotropy and polarization generated by
plane gravitational waves. After spatial Fourier transformation, the first-order Boltzmann
equation becomes:

\beq \frac{d\bf{\delta\hat{n}}(\eta,{\bf k})}{d\eta} + i k
\mu_{\bf k} {\bf{\delta\hat{n}}}(\eta,{\bf k}) = -\frac{1}{2}\gamma \frac{\partial h_{\alpha\beta}}{\partial \eta}
e^{\alpha}e^{\beta} - q(\eta) [ {\bf\delta\hat{n}}(\eta,{\bf k}) -
{\bf \hat{J}}(\eta,{\bf k})]. \label{eq:n1} \eeq where, $\mu_{\bf
k} = \frac{{{e}}^{\alpha}{{ k}_{\alpha}}}{k}$, $k = \vert {\bf
k}\vert$ and $\phi_{\bf k}$ is the azimuthal angle of
${{e}}^{\alpha}$ in the plane perpendicular to the vector ${\bf
k}$. In the case of a single gravitational wave we can choose our
spherical coordinate system in such a way that
$\cos{\theta}=\mu_{\bf k}$ and $\phi=\phi_{\bf k}$ (we shall omit
the index ${\bf k}$ in $\mu$ and $\phi$ when such an omission does
not lead to confusion).

\section{CMB polarization due to gravitational waves. An analytical approach.}

We now focus on the solutions to Eq. \ref{eq:n1} and with our attention restricted to
primordial gravitational waves (tensor perturbations). The process of generation of polarization occurs
after radiation-matter equality. In this case the source in Eq. \ref{eq:n1} has the following form:

\beq \frac{1}{2}\gamma \frac{\partial h_{\alpha\beta}}{\partial
\eta} e^{\alpha}e^{\beta} = \frac{1}{2}\gamma(1-\mu^2)\cos 2\phi
S(\eta, k), \label{eq:H}\eeq where
$S(\eta,k)=\frac{dh_k(\eta)}{d\eta}$ is the gravitational wave
source term.

For a plane gravitational wave perturbation with wavevector ${\bf
k}$, the symbolic vector ${\bf \delta\hat{n}}(\eta, {\bf k})$,
describing anisotropy and polarization, can be
%separated into a product of a time-varying term and a stationary, purely geometrical-term\cite{basko}:
presented as\cite{polnarev85}

\beq {\bf \delta\hat{n}}(\eta, {\bf k}) = \frac{\gamma}{2}\left
[\alpha(\eta,\mu ,k)(1-\mu^2)\left(
\begin{array}{c}1\\1\\0
\end{array} \right)\cos 2\phi
  + \beta(\eta,\mu, k) \left( \begin{array}{c}(1+\mu^2)\cos 2\phi \\ -(1+\mu^2)\cos 2\phi \\ 4\mu \sin 2\phi
\end{array} \right)\right ], \label{eq:pola}\eeq

For $k\eta \ll 1$, $\alpha$ and $\beta$ do not depend on $\mu$
\cite{basko}. Equation \ref{eq:pola} will allow for reconstruction
of the polarization and temperature power spectra, which can be
derived from maps of the CMB's polarization.

Substituting (\ref{eq:H}) and (\ref{eq:pola}) into the Boltzmann equation, (\ref{eq:n1}), we obtain the
following system of coupled integro-differential equations for $\alpha$
and $\beta$
\beq \dot{\beta}(\eta,\mu ,k)+(q-ik\mu)\beta(\eta,\mu ,k)=\frac{3}{16}q(\eta) I(\eta,k) \label{eq:beta},\eeq
\beq \dot{\xi}(\eta,\mu ,k)+(q-ik\mu)\xi(\eta,\mu ,k)=S(\eta,k),  \label{eq:xi}\eeq
where $\xi(\eta,\mu ,k) = \alpha(\eta,\mu ,k)+ \beta(\eta,\mu ,k)$ and
\beq I(\eta,k)=\int^{1}_{-1}d\mu{'}\left
[(1+\mu^{'2})^{2}\beta(\eta,\mu^{'},k)-\frac{1}{2}(1-\mu^{'2})^{2}\xi(\eta,\mu^{'},k)\right
].\eeq

Here $$q(\eta)=\sigma_T N_e a X_e(\eta) = \frac{\sigma_T \Omega_B
\rho_{c}X_e(\eta)}{m_p a^2},$$ where $\sigma_T$ is the Thomson
optical depth, $\Omega_B$ is the baryon fraction, $\rho_c$ is the
critical density, $m_p$ is the proton mass, $a$ is the
cosmological scale factor, and $X_e(\eta)$ is the ionization
fraction. For $X_e(\eta)$ during decoupling we use the Peebles
fitting function \cite{Peebles93}.

The formal solution of Eq. \ref{eq:beta} is \beq \beta(\eta,\mu
,k)= e^{\tau(\eta)+ik\mu\eta}\int^{\eta}_{0}\Phi(x,k)e^{-ik\mu
x}dx ,\label{beta}\eeq where $\tau(\eta)$is optical depth: \beq
\tau(\eta)=\int^{\eta_o}_{\eta}q(\eta^{'})d\eta^{'}.\eeq and \beq
\Phi(\eta,k)=\frac{3}{16}I(\eta,k)q(\eta)e^{-\tau (\eta)}. \eeq

The function $\Phi_{\eta,k}(\eta)$ depends primarily on the epoch
and duration of decoupling, and is not sensitive to the details of
$\tau(\eta)$\cite{nasel87}.

Taking into account that from Eq. \ref{eq:xi} \beq \xi(\eta,\mu
,k)=e^{\tau(\eta)+ik\mu
\eta}\int^{\eta}_{0}S(x,k)e^{-\tau(x)-ik\mu x}dx \label{xi},\eeq
 we obtain for $\Phi$ the following integral equation:
\beq
\Phi(\eta,k)=\Phi_{0}(\eta,k)+\frac{3}{16}q(\eta)\int^{\eta}_{0}\Phi(x,k)K_{+}(\eta-x,k)dx,
\label{intF} \eeq where \beq
\Phi_{0}(\eta,k)=-\frac{3}{32}q(\eta)\int^{\eta}_{0}dx
S(x,k)e^{-\tau(x)}K_{-}(\eta-x,k), \eeq and \beq
K_{\pm}(x,k)=\int^{1}_{-1}d\mu (1\pm\mu^{2})^{2}\cos k\mu x. \eeq

We remind the reader that our goal is to obtain the present value
of the symbolic vector ${\bf \delta\hat{n}}$, from Eq.
\ref{eq:pola}. To do so requires that $\Phi(\eta,k)$ be found for
every ${\bf k}$. The vector ${\bf \delta\hat{n}}$ is found by by
using equations (\ref{beta}), (\ref{xi}) and substituting into Eq.
(\ref{eq:pola}), setting $\eta=\eta_0$ and $\tau(\eta_0)=0$. Once
the symbolic vector ${\bf \delta\hat{n}}$ is found, we proceed to
determine the multipole expansion of the GWB-induced polarization,
which will eventually be used to construct statistical estimators
of observable quantities such as the temperature and polarization
power spectra.

%%%%%%%%%%%%%%%%%%%%%%%%%%%%%%%%%%%%%%%%%%%%%%%%%%%%%%%%%%%%%%%%%%%%%%%%%%%%%%%%%%
\section{Multipole expansion of Anisotropy and Polarization due to gravitational waves}

This section is largely based on Seljak \&
Zaldarriaga\cite{Zalandseljak97}. The components of the symbolic
vector ${\bf n}(\mu,\phi)$ are related to the fundamental
polarization tensor $P_{ab}(\mu,\phi)$ in the following way\bea
P_{a}^{b}&\equiv&
\frac{1}{2}\frac{c^2}{h\nu^3}\left(\begin{array}{cc}I+Q&U
\\\\U&I-Q\end{array}\right)\nonumber\\
&&\nonumber\\
&&\nonumber\\&=&
\frac{1}{2}\left(\begin{array}{cc}n_1&-\frac{1}{2}n_3
\\\\ -\frac{1}{2}n_3&n_2\end{array}\right).\ena

From the polarization tensor $P_{ab}(\mu,\phi)$ we can form three
independent scalar fields corresponding to anisotropy and $E$, $B$
modes of polarization in the following way :
\bea
T\left(\mu,\phi\right)&=&g_{ab}\left(\mu,\phi\right)
P^{ab}\left(\mu,\phi\right),\label{t}\\
E\left(\mu,\phi\right)&=&\left[g_{ab}\left(\mu,\phi\right)
P_{~c}^c\left(\mu,\phi\right)-2P_{ab}\left(\mu,\phi\right)\right]^{;a;b},\\
B\left(\mu,\phi\right)&=&\epsilon_{~d}^a\cdot
\left[g_{ab}\left(\mu,\phi\right)\cdot
P_{~c}^c\left(\mu,\phi\right)-2P_{ab}\left(\mu,\phi\right)\right]^{;b;d},\label{B}
\ena

where repeated indices imply summation, and
$$g_{ab}\left(\mu,\phi\right)=\left(\begin{array}{cc}(1-\mu^2)^{-1}&0
\\\\ 0&(1-\mu^2)\end{array}\right)$$ is
the 2-metric on a unit sphere in coordinates $(\mu,\phi)$, and
covariant derivatives ($;$) are effected with respect to this
metric.

In terms of our variables $\alpha\left(\mu,k\right)$ and
$\beta\left(\mu,k\right)$ (see (\ref{eq:pola})) the above
definitions of $T$, $E$ and $B$ (\ref{t}-\ref{B}) give\bea
T\left(\mu,\phi,k\right)&=&
\gamma\left(1-\mu^2\right)\alpha\left(\mu,k\right)\cdot\cos{2\phi},
\nonumber \ena

\bea E\left(\mu,\phi,k\right)&=& -\gamma\left(1-\mu^2\right)
\left[\left(1+\mu^2\right)\frac{d^2}{d\mu^2}+8\mu\frac{d}{d\mu}+12\right]\beta\left(\mu,k\right)\cos{2\phi},
\nonumber\ena

\bea B\left(\mu,\phi,k\right)&=& \gamma\left(1-\mu^2\right)
\left[2\mu\frac{d^2}{d\mu^2}+8\frac{d}{d\mu}\right]\beta\left(\mu,k\right)\sin{2\phi},\nonumber
\ena

The multipole expansion coefficients for anisotropy, and $E$, $B$
modes of polarization, in terms of the introduced scalars $T$,$E$
and $B$, are defined as \bea a^T_{l,m}(k)&=&\oint d \Omega
\left[\left(Y_{l,m}\left(\mu,\phi\right)\frac{}{}\right)^*\cdot
T\left(\mu,\phi,k\right)\right], \nonumber\\&&\nonumber\\
a^E_{l,m}(k)&=&\left[\frac{(l-2)!}{(l+2)!}\right]^{\frac{1}{2}}\oint
d \Omega
\left[\left(Y_{l,m}\left(\mu,\phi\right)\frac{}{}\right)^*\cdot
E\left(\mu,\phi,k\right)\right], \nonumber\\&&\nonumber\\
a^B_{l,m}(k)&=&\left[\frac{(l-2)!}{(l+2)!}\right]^{\frac{1}{2}}\oint
d \Omega
\left[\left(Y_{l,m}\left(\mu,\phi\right)\frac{}{}\right)^*\cdot
B\left(\mu,\phi,k\right)\right], \label{teb3}\ena where
$Y_{l,m}\left(\mu,\phi\right)$ are the ordinary spherical
harmonics. From (\ref{teb3}) it follows that the multipole
coefficients vanish for $m\neq\pm2$. Taking into account the form
expressions for $\alpha$ and $\beta$ from (\ref{beta}) and
(\ref{xi}), the above expressions can be rewritten as:

\bea a^T_{l,m}(k)&=&-(-i)^l\gamma\left(\delta_{2,m}
+\delta_{-2,m}\right)\sqrt{\pi(2l+1)}\int\limits_{0}^{1}d\eta~T_k(\eta)
\left[\sqrt{\frac{(l+2)!}{(l-2)!}}~\frac{j_l(\zeta)}{\zeta2}\right],
\nonumber \ena
\bea a^E_{l,m}(k)&=&-(-i)^l\gamma\left(\delta_{2,m}
+\delta_{-2,m}\right)\sqrt{\pi(2l+1)}\int\limits_{0}^{1}d\eta~\Pi_k(\eta
)
\left[\left\{\left(2-\frac{l(l-1)}{\zeta2}\right)j_l(\zeta)-\frac{2}{\zeta}j_{l-1}(\zeta)\right\}\right],
\nonumber \ena

\bea a^B_{l,m}(k)&=&-(-i)^l\gamma\left(\delta_{2,m}
-\delta_{-2,m}\right)\sqrt{\pi(2l+1)}\int\limits_{0}^{1}d\eta~\Pi_k(\eta)
\left[2\left\{-\frac{(l-1)}{\zeta}j_l(\zeta)+j_{l-1}(\zeta)\right\}\right],
\nonumber \ena

where \bea \zeta&=&k(\eta_0-\eta),\nonumber \ena

and\bea T_k(\eta)&=& S_k(\eta)e^{-\tau(\eta)}-
\Phi_k\left(\eta\right),\nonumber\\
\Pi_k(\eta)&=&\Phi_k\left(\eta\right),\nonumber\ena and
$j_l(\zeta)$ are the spherical bessel functions.

\section{Anisotropy and polarization generated by a random field of gravitational waves}

The GWB can be considered as an isotropic random superposition of
plane waves with the following correlation relationship: \beq
\langle \stackrel{s}{c}_{\bf k}^*\stackrel{s'}{c}_{\bf
k^{'}}\rangle = \langle \stackrel{s}{c}_{\bf
k}\stackrel{s'}{c}_{\bf k^{'}}^*\rangle=\delta_{ss'}\delta ({\bf
k}-{\bf k}^{'}) , ~~~ \langle \stackrel{s}{c}_{\bf
k}\stackrel{s'}{c}_{\bf k^{'}}\rangle
 = \langle \stackrel{s}{c}_{\bf
k}^* \stackrel{s'}{c}_{\bf k^{'}}^*\rangle= 0
\label{correlationrelations}\eeq where $\langle ... \rangle$ means
averaging over realizations. Below we assume power law spectrum
for Cosmological Gravitational Wave Background, i.e.
$$\left.\frac{}{}4\pi k^3 |h_k(\eta)|^2\right|_{k\eta \ll 1} \sim k^{n_T - 1},$$ with $n_T=1$
corresponding to flat (scale-invariant) Zel'dovich-Harrison
spectrum.

Let us consider two scalar fields $X(\mu,\phi)$ and
$\tilde{X}(\mu,\phi)$ (where  $X$ and $\tilde{X}$ is any pair of
the scalars $T(\mu,\phi)$, $E(\mu,\phi)$ and $B(\mu,\phi)$). Then
their cross correlation is defined as \bea
\Gamma^{X\tilde{X}}(\mu_0)&=&\frac{1}{8\pi^2}\int d\mu d\mu'd\phi
d\phi'\times \nonumber \\ \qquad &&\times\delta\left(
\mu_0-\mu\mu'-\sqrt{(1-\mu^2)(1-\mu'^2)}\cos{\left(\phi-\phi'\right)}
\right)\langle X(\mu,\phi)\tilde{X}(\mu',\phi') \rangle. \nonumber
\\ \ena Following Fourier decomposition of $X(\mu,\phi)$ \bea X(\mu,\phi) = \frac{1}{(2\pi)^{3/2}}\int~d{\bf
k}\sum_{s=1,2}\left[X_k(\mu_{{\bf k},s},\phi_{{\bf k},s})e^{i{\bf
k\cdot x}}\stackrel{s}{c}_{\bf k}+ X_k^*(\mu_{{\bf
k},s},\phi_{{\bf k},s})e^{-i{\bf k\cdot x}}\stackrel{s}{c}_{\bf
k}^*\right], \ena and the correlation relations
(\ref{correlationrelations}), the correlation function is
presentable in the form \bea \Gamma^{X\tilde{X}}(\mu_0)=4\pi\int
k^2dk ~\Gamma^{X\tilde{X}}(\mu_0,k),\label{gamma} \ena where
$\Gamma^{X\tilde{X}}(\mu_0,k)$ is the correlation function for a
single wave  \bea
\Gamma^{X\tilde{X}}(\mu_0,k)&=&\frac{1}{8\pi^2}\int
d\mu~d\mu'~d\phi~d\phi'~\nonumber\\&&
\delta\left(\mu_0-\mu\mu'-\sqrt{(1-\mu^2)(1-\mu'^2)}\cos{\left(\phi-\phi'\right)}\right)
X_k(\mu,\phi)\tilde{X}_k^*(\mu',\phi'),\nonumber\\&&\ena

The correlation function being a function of a single angle
$\theta_0 = \cos^{-1}\mu_0$, can be expanded over Legendre
polynomials to give the following correlation function and power
spectrum \bea
\Gamma^{X\tilde{X}}(\mu_0,k)=\sum_{l=0}^{\infty}C_l^{X\tilde{X}}(k)P_{l}(\mu_0)\ena
\bea &&C_l^{X\tilde{X}}(k)=\frac{2}{2l+1}\int\limits_{-1}^{1}
d\mu_0~\Gamma^{X\tilde{X}}(\mu_0,k)~P_l(\mu_0).\label{clk}\ena
Using the addition theorem for Legendre polynomials

$$P_{l}\Big[\mu\mu^{'}+\sqrt{(1-\mu^{2})(1-\mu^{'2})}\cos(\phi
-\phi^{'})\Big]=P_{l}(\mu)P_{l}(\mu^{'})+\sum_{m=0}^{l}\frac{(l-m)!}{(l+m)!}P_{l}^{m}(\mu)P_{l}^{m}(\mu^{'})\cos
m(\phi -\phi^{'}),$$ we get \bea C_l^{X\tilde{X}}(k)&=&
\frac{1}{2l+1}\frac{1}{4\pi^2}\int d\mu_0~P_{l}(\mu_0)
d\mu~d\mu'~d\phi~d\phi'~\nonumber\\&&
\delta\left(\mu_0-\mu\mu'-\sqrt{(1-\mu^2)(1-\mu'^2)}\cos\left(\phi-\phi'\right)\right)
X_k(\mu,\phi)\tilde{X}_k^{*}(\mu',\phi'),\nonumber\\&=&
\frac{1}{2l+1}\frac{1}{4\pi^2}\int
d\mu~d\mu'~d\phi~d\phi'~P_{l}\left(\frac{}{}\mu\mu'+\sqrt{(1-\mu^2)(1-\mu'^2)}\cos\left(\phi-\phi'\right)\right)
\nonumber\\&& X(\mu,\phi,k)\tilde{X}^*(\mu',\phi',k),
\nonumber\\&=& \frac{1}{2l+1}\sum_{m=-l}^{l}{x_{l,m}(k)}\cdot
\tilde{x}_{l,m}^*(k), \ena where\bea x_{l,m}(k)= \oint d\Omega
\left[\left(Y_{l,m}(\mu,\phi)\right)^*\cdot
X(\mu,\phi,k)\frac{}{}\right],\ena \bea \tilde{x}_{l,m}(k)= \oint
d\Omega \left[\left(Y_{l,m}(\mu,\phi)\right)^*\cdot
\tilde{X}(\mu,\phi,k)\frac{}{}\right],\ena

Thus we get for the power spectrum from a single gravitational
wave

\bea C^{X\tilde{X}}_l(k)&=&\frac{1}{2l+1}\sum_{m=-l}^{l}
{a^X_{l,m}} a^{\tilde{X}*}_{l,m} \ena

The angular power spectrum from a superposition of gravitational
waves is calculated from the above expression by integrating over
all the wave numbers $k$ \bea C^{X\tilde{X}}_l &=& 4\pi\int dk
~k^2 ~C^{X\tilde{X}}_l(k), \ena

The power spectra is most conveniently determined as coefficients
of a Legendre polynomial expansion with index $l$ rather than an
expansion in Fourier modes with wavenumber $k$. In a flat universe
a perturbation of comoving wavelength $k^{-1}$ at the comoving
distance of the last scattering surface (LSS) subtends an angle
$\theta \sim 1/k$. On the other hand $\ell \sim 1/\theta$ for
small $\theta$. This simple consideration shows that the main
contribution to $C_\ell^{TT}$ with a given $\ell$ comes from
$k\sim \ell$.

Figure \ref{fig:bbmode_limits} shows $C_\ell^{BB}$ calculated
using the CAMB\cite{Lewis1999} code, which is based on
CMBFAST\cite{SeljakZal1996}, compared to our analytical
predictions. The peak multipole of the spectrum
$\ell_{peak}\simeq90$ is robust to changes in the inflationary
dynamics. The amplitude of the B-mode power spectrum, is
determined by $r$ alone, and its spatial structure is determined
solely by the age of the universe at last scattering. This
single-parameter dependence makes the B-mode polarization the most
robust probe of inflation. While scalar, or mass-energy,
perturbations are amplified by gravity, the tensor-GWB is not.
Direct detection today (redshift $z = 0$) by, e.g., LIGO/LISA is
essentially impossible since, like the CMB, the energy-density of
gravitational waves dilutes (redshifts) for waves inside the
horizon at, or before, last-scattering as the universe expands.
However the GWB imprints curl-mode polarization on the CMB
\emph{at} the surface of last-scattering ($z \simeq 1100$).
Therefore, the GWB energy-density from sub-horizon scale waves at
last-scattering was at least one trillion times larger than it is
now, which motivates the use of the last-scattering surface as the
perhaps the best ``detector" of the primordial,
inflationary-generated, GWB \cite{smithpeiriscooray2006}.

Our analytic calculations based on the solution of the Boltzmann
equation provide clear insight into the underlying physics of
polarization and are in good agreement with the results of CMBFAST
and CAMB.

%%%%%%%%%%%%%%%%%%%%%%%%%%%%%%%%%%%%%%%%%%%%%%%%%%%%%%%%%%%%%%%%%%%%%%%%%%%%%%%%%
\section{Current CMB Polarization Results}\label{sec:results}

\begin{figure}[htb]
\centerline{\psfig{file=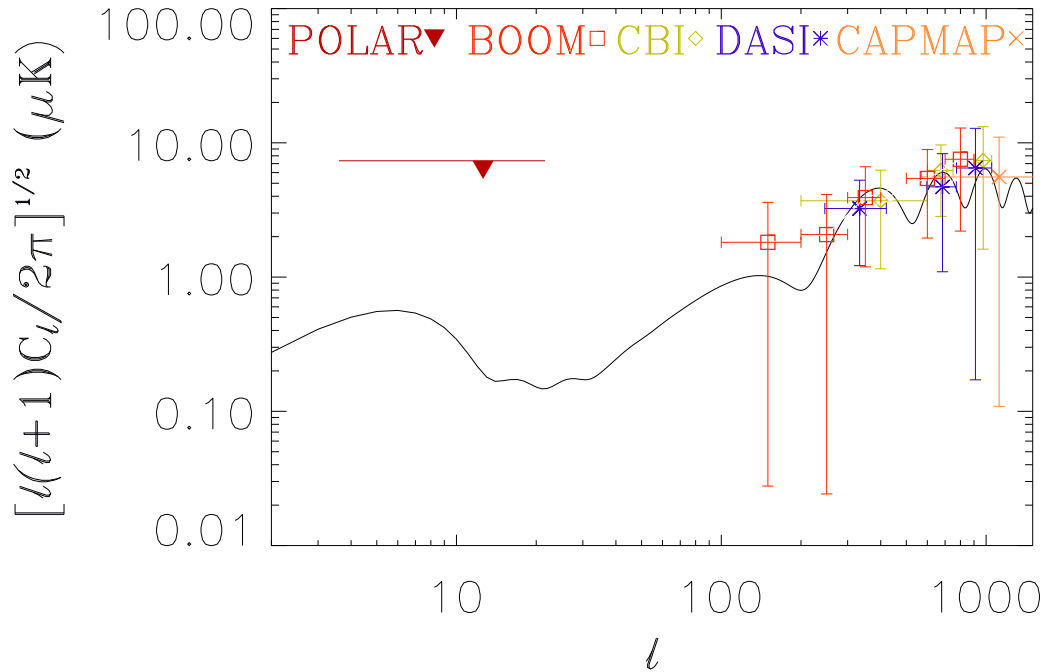, width=5.9in, angle=0}}
\caption{Measurements of the gradient or E-mode polarization power
spectrum $C_{\ell}^{E}$. The solid line is the polarization power
spectrum for the WMAP best fit cosmological model, with
$\tau=0.17$ \protect\cite{spergel2003}.} \label{fig:emode_limits}
\end{figure}

\begin{figure}[htb]
\centerline{\psfig{file=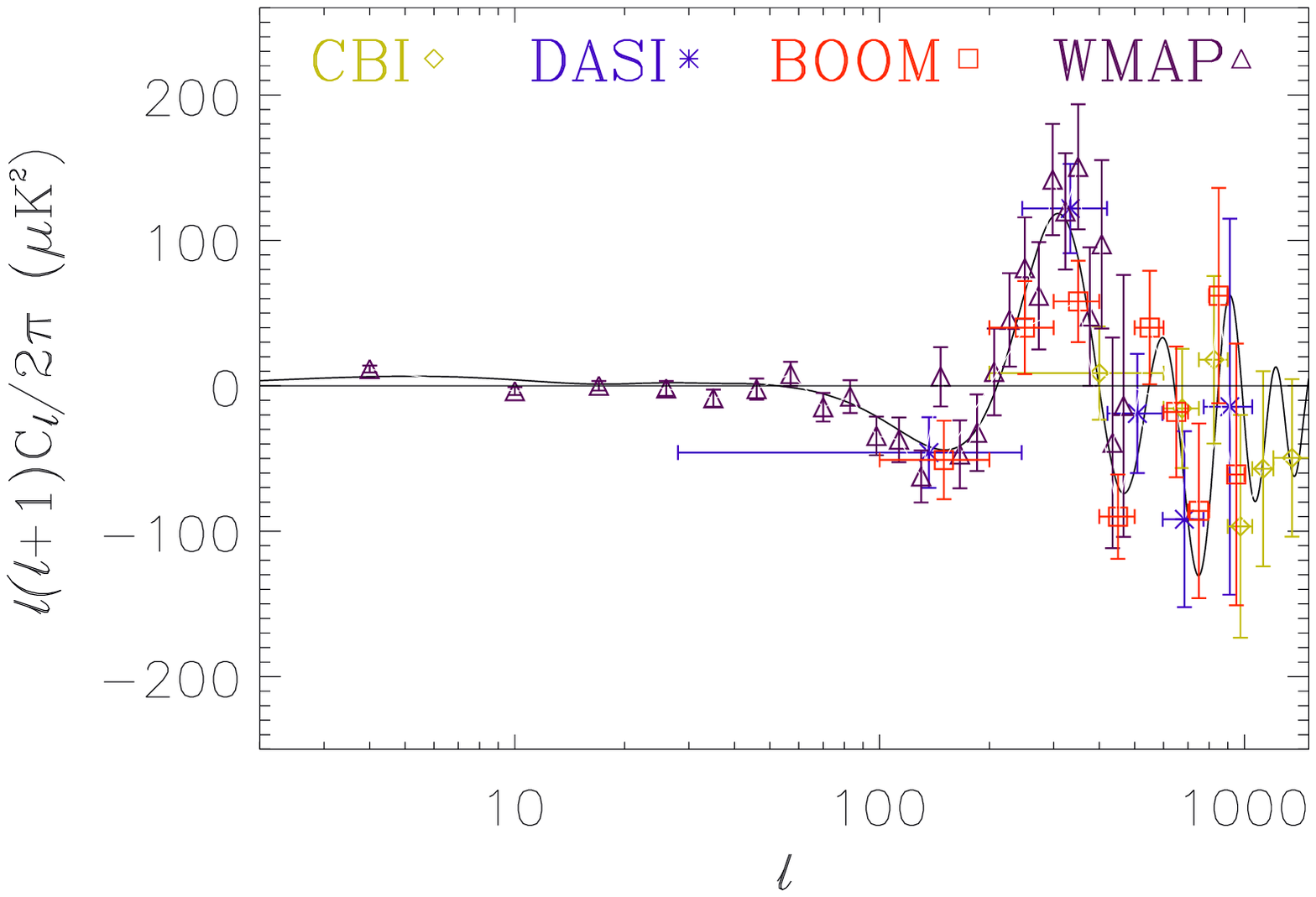, width=5.9in, angle=0}}
\caption{Measurements of the temperature-polarization
cross-correlation power spectrum $C_{\ell}^{TE}$. The solid line
is the power spectrum for the WMAP best fit cosmological model,
with $\tau=0.17$  \protect\cite{spergel2003}.}
\label{fig:temode_limits}
\end{figure}

Figures \ref{fig:emode_limits} and \ref{fig:temode_limits} show
current detections of the E-mode (grad-mode) polarization and the
polarization-temperature cross correlation, $\langle TE \rangle$.
Currently, DASI\cite{Leitch2005}, BOOMERANG\cite{Montroy05},
CBI\cite{Readhead04}, and CAPMAP\cite{Barkats2005} have detected
E-mode polarization, however there are no current detections of
the E-mode polarization for $\ell < 100$ where the signature of
gravitational waves will be manifest. Only DASI and BOOMERANG have
detected the grad-mode polarization at $100 < \ell < 400$.

Both reionization and gravitational waves imprint the polarization
of the CMB at large scales. The primary effects of reionization
are encoded in the grad-mode and temperature polarization
cross-correlation at $\ell \lesssim 50$. In the absence of
detections, the most stringent constraint in this range of
multipoles is currently $C_\ell^E < 8$ $\mu K$ at 95\% confidence
reported by POLAR\cite{Keat01} for $2 < \ell < 20$ (assuming no
B-modes). WMAP\cite{kogut2003} reports a large number of highly
significant detections, especially at low-$\ell$ due to WMAP's
ability to map the full sky. CBI, DASI, and
BOOMERANG\cite{Piacentini2005} have also detected the
cross-correlation spectrum, mainly at smaller angular scales than
WMAP. A complete description of reionization will require
detections of CMB E-mode polarization for $\ell<50$. An ancillary
benefit of reionization \cite{kaplinghatknoxsong2003} is that it
boosts the primary curl-mode power spectrum significantly near
$\ell = 10$. Due to reionization, a more stringent limit on the
tensor-to-scalar ratio, $r$, in the presence of lensing can be
obtained than that calculated in
Ref.~\refcite{knox2002,kesden2002}.

The tensor to scalar ratio, $r$, is defined here as \beq r =
\frac{\Delta_h^2(k_0)}{\Delta_R^2(k_0)} \eeq where
$\Delta_R^2(k_0)$ and $\Delta_h^2(k_0)$ are the amplitudes of the
scalar and primordial power spectra, evaluated at some pivot
wavenumber $k_0$. There are other ways of defining the tensor to
scalar ratio that appear in the literature. A second definition,
$T/S$, is defined as the ratio between the tensor and scalar
contributions to the temperature anisotropy's quadrupole component
$T/S = C^T_{2,tens}/C^T_{2,scal}$. The relationship between $r$
and $T/S$ depends in a complicated way on the cosmology. For the
parameter set used in this paper, $T/S \approx 0.5r$ where the
pivot wavenumber used is $0.05$ Mpc$^{-1}$.

CMB temperature anisotropy \emph{alone} can only detect the tensor
to scalar ratio if $r \gtrsim 0.3$ due to cosmic
variance\cite{knoxturner1994}. The limit from WMAP data alone is
roughly four-times larger than this and approximately two-times
larger with the inclusion of external data\cite{peiris2003}. Thus
we are close to the ultimate constraints achievable by CMB
temperature anisotropy observations alone.

\begin{figure}[htb]
\centerline{\psfig{file=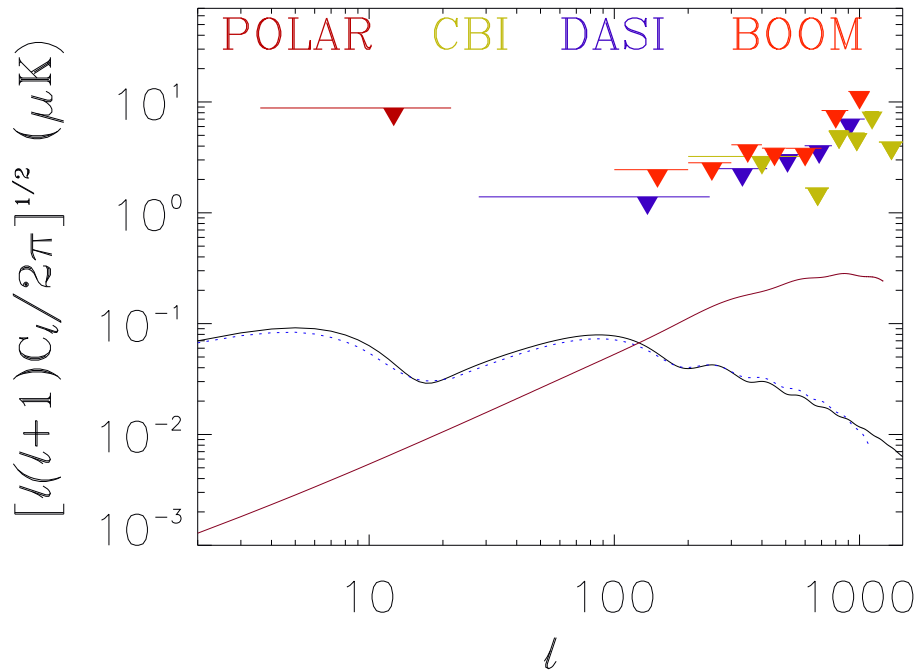, width=5.9in, angle=0}}
\caption{Measurements of the curl-mode polarization power spectrum
$C_{\ell}^{BB}$. The solid line is the power spectrum for the WMAP
best fit cosmological model with $\tau=0.17$
\protect\cite{spergel2003} and $r=0.1$. The dotted line are the
analytical results of this paper, and the solid curve which peaks
at $\ell \sim 1000$, is the B-mode spectrum produced by large
scale structure lensing the primary CMB E-mode polarization.}
\label{fig:bbmode_limits}
\end{figure}

Further progress requires detection of the B-mode signal.
Currently, there are no detections of the B-mode polarization as
shown in figure \ref{fig:bbmode_limits}. POLAR, CBI, DASI, and
BOOMERANG all provide upper limits to the B-mode polarization. The
current generation of CMB polarimeters should be able to provide
detect or provide much better upper limits to the curl-mode
polarization within the next several years.

The gravitational lensing of the E-mode polarization into B-mode
polarization provides a source of contamination to the primordial
B-mode signal. For a noise free, full sky experiment in which the
lensing is treated as noise, Ref.~\refcite{knox2002} found that
this contamination sets a detectability limit of $r_{lim} >
10^{-4}$,\cite{knox2002,kesden2002} i.e. if the energy scale of
inflation is larger than $3 \times 10^{15} GeV$. Measurements of
$21$-cm radiation may be able to provide a way to de-lens the
curl-mode measurements, providing an detectable inflationary
energy scale limit lower than one using CMB-only measurements.
This energy scale may be as low as $3 \times 10^{14}$
GeV\cite{SigurdsonCooray05}. Work by Seljak \& Hirata (2004)
\cite{seljak2004} indicate that the primary lensing signal can be
removed to a level that makes $r=10^{-6}$ detectable, allowing
detection of inflationary GWB at energy scales $< 10^{15}$ GeV
using the CMB only.

\section{Constraining Inflation with Upcoming CMB Observations
\label{section:inflation_predictions}} 
\begin{figure}[htb]
\centerline{\psfig{file=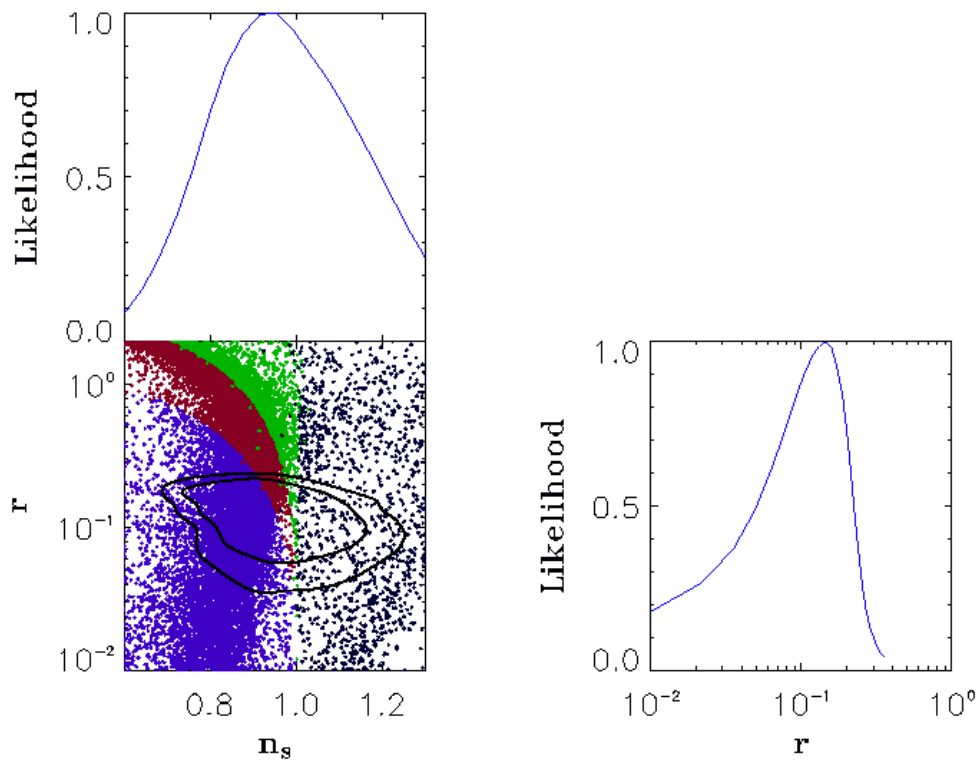, width=5.9in, angle=0}}
\caption{Shown are 1D marginalized probabilities and 2D joint
likelihood contours for $n_s$ and $r$ from the toy experiment
described in the text. The purple dots are negative curvature
models. The red dots are small positive curvature models. The
green dots are intermediate positive curvature models. The black
dots are large positive curvature models. WMAP has set an upper
limit on $r$ of $r<1.28$ with no priors. The toy experiment is
capable of detecting $r=0.12$ with 95\% confidence with no
priors-- more than an order of magnitude below the WMAP limits.}
\label{fig:nsr_limits}
\end{figure}

To illustrate the power of
upcoming observations to constrain inflationary parameters we
consider a hypothetical polarimeter with $1^{\circ}$ resolution
and a system sensitivity ($NET_{sys}$) $= 70\muK s^{1/2}$ using
the COSMOMC\cite{Lewis02} software package. The system sensitivity
for polarization is $\sqrt{2}$ times highe, $NEQ = \sqrt{2}NET$.
These detector requirements are well within the reach of the
current generation of CMB polarimeters. By observing $2.4\%$ of
the sky, this experiment could feasibly detect the curl mode
signal created by the gravitational wave background (at $\ell <
100$) for $r$ as low as $r=0.12$ at $95\%$ confidence with no
priors. This represents an order of magnitude improvement over the
WMAP only results of $r = 1.28$ at 95\% confidence, with no
priors. We plot the results of our simulations along with
predictions for several classes of inflationary models in figure
\ref{fig:nsr_limits}. The predictions of the various models of
inflation were generated using the method described below.

Detecting the tensor-to-scalar ratio, $r$, will allow the
possibility of distinguishing between several classes of
inflation, such as negative curvature models, small positive
curvature models, etc. As an example we consider the toy
experiment described earlier. As mentioned, it can detect $r$ with
$95\%$ at $0.12$. The $n_s-r$ contours compared to the different
classes of inflation models are plotted in Figure
\ref{fig:nsr_limits}, assuming the toy experiment detects $r=0.12$
and $n_s=0.98$. The contours for the toy experiment were
calculated using COSMOMC.

The predictions for the different classes of inflation were
generated following the model in
Ref.~\refcite{kinney02,esthkinney02} and the different inflation
models were separated according to Ref.~\refcite{peiris2003}. The
observables $r$ and $n_s$ were evaluated at some specific
\textit{e}-folding, $N$, of inflation, not a specific wavenumber,
$k$. The relation between $N$ and $k$ requires a detailed model of
reheating, which has some uncertainty. This uncertainty is
marginalized over by calculating the observables at an
\textit{e}-fold randomly drawn from $40$ to $70$. The inflationary
flow equations were truncated at sixth order and the observables
were calculated to second order in slow roll. Each \emph{class} of
model has a unique color in Fig. \ref{fig:nsr_limits} as described
in the caption, and each \emph{realization} of that model is
indicated by a point in the $r-n_s$ plane.

\section{Conclusions}

We have developed an analytic method to generate the predictions
of the imprint of gravitational waves on the CMB. Using
phenomenological models of inflation we predict both the CMB
polarization spectra and the derived inflationary parameters: the
tensor-to-scalar ratio and spectral index of the scalar
perturbations. The combination of the later two observables allows
for reconstruction of the dynamics of inflation. With these
predictions in hand we have shown that upcoming CMB polarization
observations will be able to detect or constrain the cosmological
GWB and hence, inflation itself. These new technological advances
now position observational cosmology at the threshold of an
exhilarating era -- one in which CMB polarization data will winnow
down inflation's vast model-space and test models of the early
universe at energy scales approaching the GUT-scale; nearly one
trillion times higher energy than accessible from particle
accelerators.

\section*{Acknowledgments}
BK gratefully acknowledges many helpful comments on this
manuscript from Meir Shimon-Moshe and Asantha Cooray. Use of the
San Diego Supercomputing Center is gratefully acknowledged. NM is
supported by a fellowship from the U.S. Department of Education's
GAANN program. This work was supported in part by NSF CAREER Award
\#0548262.

\bibliography{Inflation,Polarization}
\bibliographystyle{unsrt}
%\begin{thebibliography}{0}
%\end{thebibliography}

\end{document}